\documentclass[conference]{IEEEtran}

\usepackage{silence}
\usepackage[font=small, labelfont=bf]{caption}

\IEEEoverridecommandlockouts

\usepackage{cite}
\usepackage{amsmath,amssymb,amsfonts}
\usepackage{algorithmic}
\usepackage{graphicx}
\usepackage{textcomp}
\usepackage{xcolor}
\def\BibTeX{{\rm B\kern-.05em{\sc i\kern-.025em b}\kern-.08em
    T\kern-.1667em\lower.7ex\hbox{E}\kern-.125emX}}

\usepackage{soul}
\usepackage{booktabs}
\usepackage{multirow}
\usepackage{graphicx}
\usepackage{enumitem}

\usepackage{tcolorbox}

\usepackage{mdframed}

\usepackage{color}
\usepackage{xcolor}
\usepackage{listings}
\usepackage{float}   

\definecolor{codegreen}{rgb}{0,0.6,0}
\definecolor{codegray}{rgb}{0.5,0.5,0.5}
\definecolor{codepurple}{rgb}{0.58,0,0.82}
\definecolor{backcolour}{rgb}{0.95,0.95,0.92}

\lstdefinestyle{mystyle}{
    backgroundcolor=\color{backcolour},   
    commentstyle=\color{codegreen},
    keywordstyle=\color{magenta}\bfseries,
    moredelim=[is][\color{magenta}\bfseries]{@}{@},
    numberstyle=\tiny\color{codegray},
    stringstyle=\color{codepurple},
    basicstyle=\C\footnotesize,
    breakatwhitespace=false,         
    breaklines=true,                 
    captionpos=b,                    
    keepspaces=true,                 
    numbers=left,                    
    numbersep=5pt,                  
    showspaces=false,                
    showstringspaces=false,
    showtabs=false,                  
    tabsize=2
}

\lstdefinestyle{Update_mystyle}{
    backgroundcolor=\color{backcolour},   
    commentstyle=\color{codegreen},
    keywordstyle=\color{magenta}\bfseries,
    moredelim=[is][\color{magenta}\bfseries]{@}{@},
    numberstyle=\tiny\color{codegray},
    stringstyle=\color{codepurple},
    basicstyle=\C\footnotesize,
    breakatwhitespace=false,         
    breaklines=true,                 
    captionpos=b,                    
    keepspaces=true,                 
    numbersep=5pt,                  
    showspaces=false,                
    showstringspaces=false,
    showtabs=false,                  
    tabsize=2
}

\lstdefinestyle{custom}{
    basicstyle=\tiny\ttfamily, 
    backgroundcolor=\color{white},          
    frame=lines,              
    framesep=4pt,             
    xleftmargin=4pt,          
    xrightmargin=4pt,          
    numbers=none              
}

\usepackage{url}

\begin{document}



\title{Enabling High-Throughput Parallel I/O in Particle-in-Cell Monte Carlo Simulations with openPMD and Darshan I/O Monitoring 
}

\author{
\IEEEauthorblockN{Jeremy J. Williams$^1$, Daniel Medeiros$^1$, Stefan Costea$^2$, David Tskhakaya$^3$, Franz Poeschel$^4$,  René Widera$^4$, \\ Axel Huebl$^5$, Scott Klasky$^6$, Norbert Podhorszki$^6$, Leon Kos$^2$, Ales Podolnik$^3$, Jakub Hromadka$^3$, \\  Tapish Narwal$^4$, Klaus Steiniger$^4$, Michael Bussmann$^4$,  Erwin Laure$^7$, Stefano Markidis$^1$}
\IEEEauthorblockA{\textit{$^1$KTH Royal Institute of Technology, Sweden}; \textit{$^2$LeCAD, University of Ljubljana, Slovenia}; \\ \textit{$^3$Institute of Plasma Physics of the CAS, Czech Republic}; \textit{$^4$Helmholtz-Zentrum Dresden-Rossendorf, Germany}; \\ \textit{$^5$Lawrence Berkeley National Laboratory, USA}; \textit{$^6$Oak Ridge National Laboratory, USA} \\ \textit{$^7$Max Planck Computing and Data Facility, Germany} 
}}

\maketitle


\begin{abstract}
Large-scale HPC simulations of plasma dynamics in fusion devices require efficient parallel I/O to avoid slowing down the simulation and to enable the post-processing of critical information. Such complex simulations lacking parallel I/O capabilities may encounter performance bottlenecks, hindering their effectiveness in data-intensive computing tasks. In this work, we focus on introducing and enhancing the efficiency of parallel I/O operations in Particle-in-Cell Monte Carlo simulations. We first evaluate the scalability of BIT1, a massively-parallel electrostatic PIC MC code, determining its initial write throughput capabilities and performance bottlenecks using an HPC I/O performance monitoring tool, Darshan. We design and develop an adaptor to the openPMD I/O interface that allows us to stream PIC particle and field information to I/O using the BP4 backend, aggressively optimized for I/O efficiency, including the highly efficient ADIOS2 interface. Next, we explore advanced optimization techniques such as data compression, aggregation, and Lustre file striping, achieving write throughput improvements while enhancing data storage efficiency. Finally, we analyze the enhanced high-throughput parallel I/O and storage capabilities achieved through the integration of openPMD with rapid metadata extraction in BP4 format. Our study demonstrates that the integration of openPMD and advanced I/O optimizations significantly enhances BIT1's I/O performance and storage capabilities, successfully introducing high throughput parallel I/O and surpassing the capabilities of traditional file I/O.
\end{abstract}





\section{Introduction}
Large-scale plasma simulations on high-performance systems continue to revolutionize our understanding of complex physical phenomena, particularly in the realm of fusion energy research. These simulations enable scientists to delve into the dynamics of plasma, crucial for optimizing the performance and safety of fusion reactors. From investigating turbulence and instabilities to probing confinement properties, these simulations offer invaluable insights that pave the way for advancements in sustainable energy production. However, the efficiency of these simulations heavily relies on efficient data handling and management, making high-performance I/O systems indispensable. These systems not only facilitate the seamless flow of data during simulations but also play a pivotal role in ensuring instantaneous post-processing of critical information. 

Introducing parallel I/O in Particle-in-Cell (PIC) Monte Carlo (MC) simulations is particularly significant as it enables the efficient handling of data streams from multiple computational processes concurrently. This parallel approach enhances throughput, reducing the time required for data storage and retrieval, thereby accelerating the pace of scientific discovery, and expanding the scope of simulations. Despite the remarkable strides in computing capabilities, the persistent challenge lies in mitigating the performance bottleneck posed by I/O systems, which can impede the pace of scientific discovery and limit the scope of PIC MC simulations.

The openPMD standard~\cite{openPMDstandard} aims to solve the issue of portability of exchange particle and mesh based data from scientific simulations and experiments by providing a minimal set of meta information, supporting mutiple backends such as HDF5, ADIOS1, ADIOS2 and JSON in both serial or MPI-based workflows. BIT1 is a parallel plasma simulation electrostatic PIC MC code that currently contains I/O bottlenecks, and currently do not implement the openPMD standard. 

In this work, we focus on introducing and enhancing throughput parallel I/O in BIT1, achieved through assessing and monitoring traditional file I/O performance using the Darshan tool, integrating the openPMD standard, facilitating efficient I/O operations, supporting parallel workflows, and minimizing data and storage on the file system. The contributions of this work include: 

\begin{itemize}[leftmargin=*]
    \item We evaluate the performance of traditional file I/O in BIT1 with diagnostics activated.
    \item We critically discuss how the usage of a standard for naming schema can benefit a plasma simulation application.
    \item We implement an I/O adaptor for the openPMD I/O interface that uses ADIOS2 BP4 as I/O interface.
    \item We show that our high-throughput parallel I/O approach offers significant benefits on the POSIX interface.
    \item We analyze the impact of varying the number of aggregators, using data compressors, and explore optimization strategies for the BIT1 storage, focusing on Lustre file striping parameters such as stripe counts and sizes.
\end{itemize}

The remainder of this paper is organized as follows. Section~2 provides background information on ADIOS Version 2, openPMD Standard \& openPMD-api, and BIT1's I/O strategies. Section~3 details our methodology and the experimental setup, including modifications made to BIT1 for this work. Performance results are presented in Section~4, including benchmarking, I/O costs per process analysis, data compression, aggregation, and Lustre file striping. Related work is discussed in Section~5. Finally, Section~6 contains the discussion, conclusion and future work.


\section{Background}\label{sec:bg}
The PIC method is a numerical approach used to model plasma behavior. The method simulates particle dynamics in one to three-dimensions (1-3D) in a usual space and typically three dimensions (3V) in the velocity space. For plasma edge applications, the PIC method is usually complemented by MC routines for simulation of particle collisions and their interaction with the plasma device chamber walls. There are five phases to the computational PIC cycle: plasma density calculation using particle-to-grid interpolation, a density smoothing process to eliminate spurious frequencies, a field solver solving a linear system for electric and magnetic fields, addressing particle collisions and wall interactions with a MC technique, and advancing particle positions and velocities through time.



The application tool used in this work, BIT1, has a workflow used to simulate the plasma edge~\cite{tskhakaya2004magnetised, tskhakaya2007optimization, tskhakaya2010pic}, plasma behavior in the tokamak divertor  and is designed for large-scale plasma simulations on high-performance computing (HPC) systems. BIT1 is a 1D3V electrostatic PIC MC code used for plasma, impurity and neutral transport in a 1D magnetic flux tubes of the magnetic confinement fusion plasma edge. Despite low dimensional BIT1 can capture large number of kinetic processes and enable corresponding pioneering studies~\cite{tskhakaya2017one}. The Input to BIT1 represents a relatively small (1-3 kB) file read by all processes. The output corresponds to three different processes defined by five critical input parameters:

\begin{itemize}[leftmargin=*]
    \item \textbf{datfile}: Captures the system's diagnostic snapshot at a specific time step.
    \item \textbf{dmpstep}: Determines when the simulated system's current state is saved, indicating the preservation of particle states or time steps.
    \item \textbf{mvflag}: Represents a flag for activating time-dependant diagnostics of plasma profiles and particle angular, velocity and energy distribution functions. If $>$ 0 it determines the number of time steps steps at which time dependent diagnostics are averaged.
    \item \textbf{mvStep}: Counter time steps for the interval between the time-dependant diagnostics. 
    \item \textbf{last\_step}: Marks the time step at which the code concludes, saving the present state on the disk and terminating the simulation.
\end{itemize}

While the original version of BIT1's serial output functioned well for runs using up to 20,000 MPI Processes, larger simulations presented challenges. Beyond this threshold, the output process demanded considerably more time and frequently resulted in corrupted files. To ensure the accuracy of output files and optimize performance for extensive simulations, novel parallel I/O methods will need to be implemented in BIT1.  

\subsection{ADIOS Version 2}
The Adaptable Input Output System version 2 (ADIOS2) is an open-source framework designed for scalable parallel I/O. It supports the usage of C, C++, Fortran, and Python across various device platforms, including supercomputers, personal computers, and cloud systems. Its unified application programming interface (API) emphasizes n-dimensional variables, attributes, and steps, abstracting low-level details for efficient data transportation in applications such as checkpoint-restart storage, code-coupling data streaming, and in situ analysis and visualization workflows~\cite{pugmire2022adaptable}.


One of the most relevant features of ADIOS2 is the ability to choose the processing virtual engine. These engines relates to how the data will be read and written, and also gives the users different parameters to be set. These engines are the BP5, BP4, BP3, HDF5, Sustainable Staging Transport, Strong Staging Coupler, the DataMan, and the Inline engines. This work explores the usage of the BP4 engine. This is because BP4 prioritizes I/O efficiency at a large scale through aggressive optimization, while BP5 incorporates certain compromises to exert tighter control over the host memory usage. 

\subsection{openPMD Standard \& openPMD-api}

The openPMD standard~\cite{openPMDstandard}, known as the open standard for particle-mesh data files, provides portability for exchanging particle and mesh-based data from scientific simulations. It offers minimal meta information and supports diverse backends, including HDF5, ADIOS1, ADIOS2, and JSON, in both serial and MPI-based workflows. The openPMD-api library serves as a reference API for openPMD data handling~\cite{openPMDapi,openPMDstandard} and scientific I/O, as depicted in~\cite{poeschel2021transitioning}, supporting various backends such as HDF5, ADIOS2, and JSON. It accommodates both serial and MPI-parallel workflows, facilitating seamless writing and reading across different file formats.

In openPMD, a record is a physical quantity of arbitrary dimensionality (rank), potentially with multiple record components (e.g., scalars, vectors, tensors). These records share common properties, e.g., describing an electric or density field or particle property. Records may be structured as meshes (n-dimensional arrays) or not, the latter case being the storage of particle species in 1D arrays, where each row represents a particle. 

Updates to the values of the meshes or particle species is named as an iteration, and can be used to store the evolution of records over the time. Finally, the collection of iterations are named series, in which openPMD-api focuses on implementing many backends and encoding strategies.

\subsection{Fundamental Library and Interface} 
The Standard Input/Output (stdio) library is a fundamental component in many programming languages, including C and C++, and designed for handling interactions with files or streams (through functions such as \texttt{fopen}, \texttt{fread}, \texttt{fprintf}) and to the ``stdout". This library plays a significant role in basic I/O operations within the BIT1 code: when it is used within BIT1, it normally handles tasks as reading user input for simulation parameters, providing progress updates during the simulation execution, and basic logging. However, the usage of stdio in extreme-scale PIC MC simulation codes might not be the most efficient solution for extreme-scale PIC MC Simulation codes. 

Meanwhile, the Portable Operating System Interface (POSIX) defines interface between an operating system and application programs, providing compatibility among operating systems, which is often related to file consistency guarantees. It includes a set of APIs for tasks such as file I/O, process management, and inter-process communication. POSIX threads could be used to parallelized certain aspects of the code, improving its performance on multi-core systems and may be leveraged for handling file-related operations in a portable and standardized manner. Many flavours of MPI (used by ADIOS2 BP engines) implementations are POSIX-neutral, which means that it is compatible with most operating systems.


\section{Methodology \& Experimental Setup} \label{sec:method}


Building upon our focus outlined in this work, which centers on integrating the openPMD standard, facilitating efficient I/O operations, supporting parallel workflows, and minimizing data storage, we detail the specific modifications implemented within BIT1. 

\subsection{openPMD \& openPMD-api Integration}
BIT1 is a 1D3V PIC code implemented in C, where simulations are executed in one-dimensional space with the utilization of three dimensions for particle velocities. The openPMD-api parallel I/O library has been seamlessly integrated into BIT1 to enhance its I/O functionality. 

In~\cite{bit1openpmdbp4}, a new header file, \texttt{bit1.hpp}, written in C++, has been created to integrate openPMD into BIT1, providing accommodation for the functions and global variables associated with the openPMD standard. In this header file, key components include the combination of the original BIT1 header, originally written in C, a vital "Series" object acting as the root of the openPMD output, extending across all data for all iterations~\cite{example1structure}. This object is made universally accessible throughout the code. Additionally, a collection of vectors, whether singular or nested, of type \texttt{float} or \texttt{unsigned long} is used to store data until either \texttt{flush()} is invoked or until the "iteration" is formally concluded. It is important to note that once an ``iteration" is closed, reopening it is not required~\cite{exampleC++writing}. Accompanying these structures are dedicated functions for converting data arrays into openPMD data objects and for the subsequent storage of these objects onto the disk. 

\subsection{Checkpointing \& Compressors}
BIT1 operates with minimal diagnostics, as evidenced by the time history of the total particle number~\cite{kovcan2006particle}. Additionally, depending on the selected options in the input file, it can log particle and power fluxes to the wall with minor computational overhead. Furthermore, BIT1 has the capability to periodically save the system's state for checkpointing and restart purposes.

To enhance data transmission speed and reduce the volume of large datasets within BIT1 simulations, Blosc data compression has been integrated~\cite{blosc2compressor,zeyen2017cosmological}, which can be compared to the high-quality data compressor bzip2~\cite{seward1996bzip2,julian2000bzip2}. This implementation utilizes a ``TOML-based dynamic configuration" with a ``group-based iteration encoding with steps" memory strategy~\cite{group-based_iteration_encoding_with_steps}, as shown in~\cite{bit1openpmdbp4}. 

\begin{figure*}[ht]
  \begin{center}
  \includegraphics[width=0.95\textwidth]{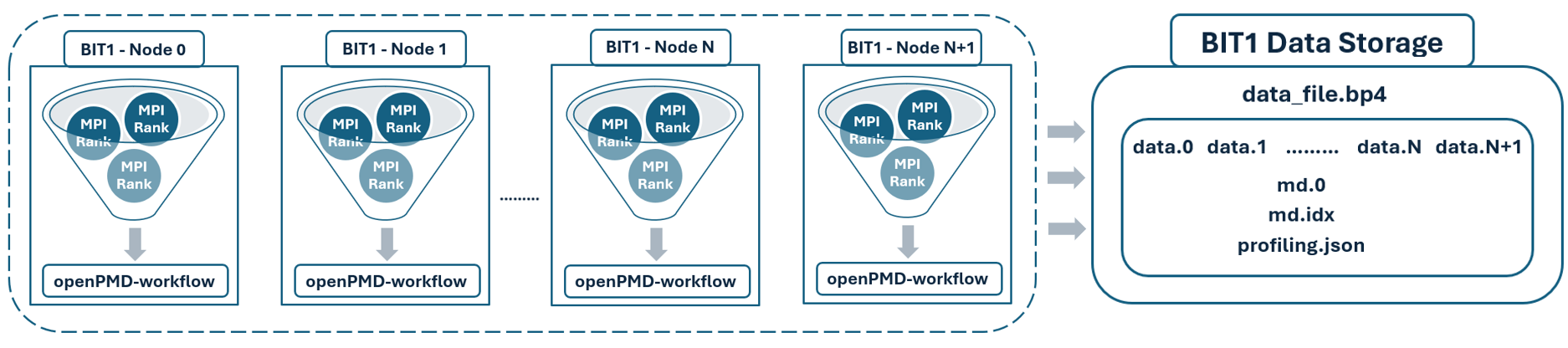} 
  \caption{BIT1 I/O Workflow with openPMD using ADIOS2 BP4 Engine}
  \label{motivation_openPMD_workflow}
  \end{center}
\end{figure*}

To use these functionalities using openPMD, all MPI ranks must adhere to a step-by-step procedure, unless specified otherwise. First, the Blosc compressor configuration is passed (or not) to the constructor of the Series class, then the file must be opened for writing to disk by creating a ``Series" object with the filename's path, access mode, and the global MPI communicator as arguments. The file's extension dictates the engine used by openPMD for data storage. Subsequently, the iteration holding the data must be explicitly opened. 

For BIT1, iteration 0 is chosen to record data that is periodically overwritten, such as the latest system state for simulation continuation. Following this, any function that stores data with openPMD must be called, with each MPI rank creating a local vector to store values for subsequent saving. These local vectors are then appended to global vectors, ensuring data persistence until \texttt{flush()} is called. Once data accumulation is complete, the accumulated data is flushed to disk in a single action for optimal I/O efficiency. The iteration is then explicitly closed, and if no further iterations are needed, the series is closed, and all global vectors are cleared.

When BIT1 utilizes openPMD, the contents of the function \texttt{any\_function\_save()} involve creating vectors locally by each MPI rank to store values intended for saving. These local vectors are then appended to global vectors and subsequently cleared. The length of the local vector is referred to as the local extent. To save to the disk, openPMD requires both the local extent and the offset of each MPI rank in the global extent of the array to be saved, and this information is obtained by calling MPI functions. Additionally, openPMD-specific objects such as ``Iteration", into which data is saved, ``Datatype" of the vector to be saved to disk, ``Global Extent", ``Offset" and ``Local Extent" objects, ``Dataset", and ``Record Component" are created to facilitate the required compiler linking process. Finally, if the local vector is not empty, it is stored to disk.

It is crucial to note that key operations between \texttt{storeChunk()} and \texttt{flush()} must not modify the referenced data, as emphasized throughout the openPMD documentation. This integrated approach enhances the capabilities of BIT1 and facilitates efficient data storage and retrieval, ensuring robust checkpointing and restoration mechanisms.

\subsection{Use Case \& Experimental Environment}

In this work, we focus on enhancing both scalability and I/O efficiency in BIT1 using openPMD and ADIOS2 backends. We simulate neutral particle ionization resulting from interactions with electrons in upcoming magnetic confinement fusion devices like ITER and DEMO. The scenario involves an unbounded unmagnetized plasma consisting of electrons, $D^+$ ions and $D$ neutrals. Due to ionization, neutral concentration decreases with time according to $\partial n / \partial t = n n_e R$, where $n$, $n_e$ and $R$ are neutral particles, plasma densities and ionization rate coefficient, respectively. We use a one-dimensional geometry with 100K cells, three plasma species ($e$ electrons, $D^+$ ions and $D$ neutrals), and 10M particles per cell per species. The total number of particles in the system is 30M. Unless differently specified, we simulate up to 200K time steps. An important point of this test is that it does not use the Field solver and smoother phases (shown in \cite{williams2023leveraging,williams2024optimizing}).

We simulate and evaluate the I/O performance of BIT1 on the following three distinct systems:

 \begin{itemize}[leftmargin=*]
\item \textbf{Discoverer}, a petascale EuroHPC supercomputer, features a CPU partition with 1128 compute nodes. Each node is equipped with two AMD EPYC 7H12 64-Core processors, 256 GB DDR4 SDRAM (on regular nodes), 1TB DDR4 SDRAM (on fat nodes), interconnected using Ethernet Controller I350 with 10 GiB/s Bandwidth and Mellanox ConnectX-6 InfiniBand with the Dragonfly+ topology, amounting to 200 GiB/s Bandwidth. For storage, \textbf{Discoverer} has a Network File System (over Ethernet) with 4.4 TB, and a Lustre File System (LFS) with 2.1 PB in capacity and 4 Object Storage Targets (OSTs). The operating system (OS) is Red Hat Enterprise Linux release 8.4, and all the applications were compiled with GCC 11.4.0 and MPI library, MPICH 4.1.2 for intra-node communication.

\item \textbf{Dardel}, an HPE Cray EX supercomputer, features a CPU partition with 1270 compute nodes. Each node is equipped with two AMD EPYC™ Zen2 2.25 GHz 64-core processors, 256 GB DRAM, and interconnected using an HPE Slingshot network with the Dragonfly topology, amounting to 200 GiB/s Bandwidth. In terms of storage, Dardel has a LFS with 12 PB in capacity and 48 OSTs. The OS is SUSE Linux Enterprise Server 15 SP3, and all applications were compiled with GCC 11.2, openPMD 0.15.2, ADIOS2 2.10.0 (with Blosc and bzip2 compression enabled) and Cray MPICH 8.1 as the MPI flavor for intra-node communication.

\item \textbf{Vega}, a petascale EuroHPC supercomputer, features a CPU partition with 960 compute nodes. Each node is equipped with two AMD EPYC 7H12 64-Core processors, 256 GB DDR4 SDRAM (80\%/nodes), 1TB DDR4 SDRAM (20\%/nodes), interconnected using Mellanox ConnectX-6 InfiniBand HDR100 with a Dragonfly+ topology, amounting up to 500 GiB/s Bandwidth. For storage, \textbf{Vega} has a Ceph File System (CephFS) with 23 PB, and LFS with 1 PB in capacity and 80 OSTs. The OS is Red Hat Enterprise Linux 8, and all applications were compiled with GCC 12.3.0 and MPI library, OpenMPI 4.1.2.1 for intra-node communication.
\end{itemize}



\subsection{I/O Workflow \& Monitoring} 
As pointed by~\cite{williams2023leveraging}, BIT1 performs serial I/O during every simulation. Similar to~\cite{poeschel2021transitioning}, a workflow has been implemented by utilizing the selected ADIOS2 engines with the required output extensions (\texttt{.bp}, \texttt{.bp4}, and \texttt{.bp5} respectively). For each extension, a unique ADIOS2 file (or directory) is created right after each simulation run, containing one or more data files (\texttt{data.0, data.1 ... data.N, data.N+1}), a metadata file (\texttt{md.0}), an index table (\texttt{md.idx}), and if enabled, a profiling file (\texttt{profiling.json}). However, for BP5, there is a second metadata file (\texttt{mmd.0}) in the directory, which BP4 and BP3 do not have. Figure~\ref{motivation_openPMD_workflow} displays the BIT1 I/O Workflow with openPMD using ADIOS2 BP4 Engine, output extension directory \texttt{data\_file.bp4} and default profiling enabled.


Among the many I/O profilers, Darshan stands out as a tool of choice to provide insight into how applications interact with underlying storage systems. Darshan is a performance monitoring tool specifically designed for analyzing both serial and parallel I/O workloads~\cite{snyder2016modular}. It collects various statistics during runtime, including data transfer sizes, access patterns, and file metadata operations. These metrics are relevant for understanding I/O bottlenecks (further discovered in~\cite{williams2023leveraging}) and fine-tuning application performance. We evaluate the I/O performance of BIT1 in terms of write throughput by extracting the throughput and amount of data stored by each file on the file system using Darshan 3.4.2 logs.


In our tests, we initially evaluate BIT1 with a standard I/O setup (BIT1 Original I/O) to assess the impact of scaling the number of nodes on Discoverer, Dardel, and Vega CPU LFS. Next, focusing on Dardel's LFS, we use different BIT1 configurations and compare BIT1 Original I/O with others using openPMD and ADIOS2 engine (BP4), aggregators (1 to 25600), data compressors (Blosc, bzip2), Lustre stripe counts (1, 2, 4, 8, 16, 32, 48), Lustre stripe sizes (1MB, 2MB, 4MB, 8MB, 16MB), and the number of parallel writers.


\section{Performance Results}\label{sec:res}

We begin this work by investigating the I/O performance when BIT1 diagnostics are activated. To better understand scalable I/O performance, we use a simulation of 200K time steps, and we have diagnostics output (with BIT1 I/O flags \texttt{slow} for plasma profiles and distribution functions, \texttt{slow1} for self-consistent atomic collision diagnostics, generating the required \texttt{.dat} files) every 1K cycles and checkpointing files (so-called \texttt{.dmp} files) every 10K cycles. The read operations are limited to read the simulation input files. In this work, we focus on analysing the performance of the write operations in BIT1. 

\begin{figure}[htbp]
    \begin{center}
        \includegraphics[width=0.95\linewidth]{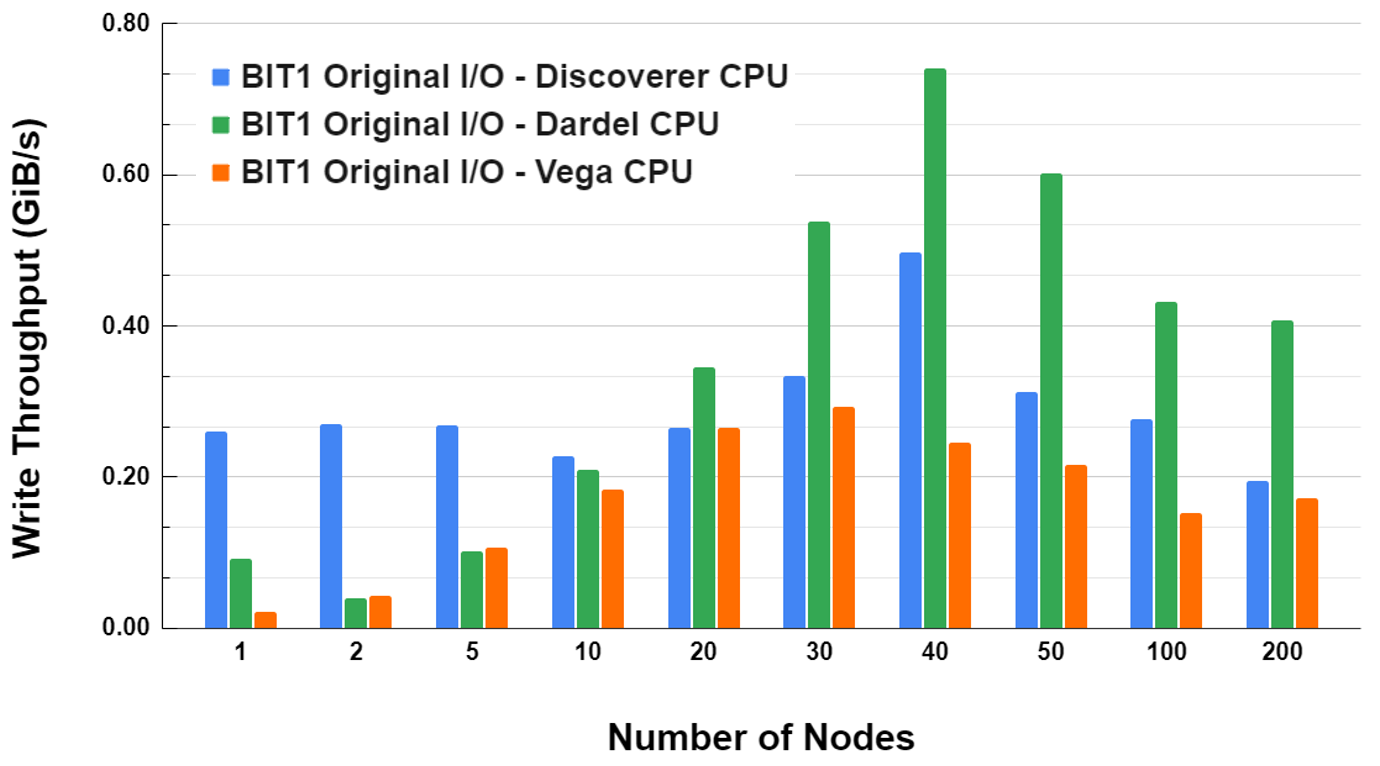}
        \caption{BIT1 Original File I/O Write Throughput, on Discoverer, Dardel and Vega CPU LFS, up to 200 Nodes, measured in GiB/s.} \label{darshan_BIT1_Original_IO}
    \end{center}
\end{figure}

Fig.~\ref{darshan_BIT1_Original_IO} displays the performance of traditional file I/O in BIT1 on Discoverer, Dardel, and Vega CPU LFS. The Discoverer CPU exhibits fluctuating performance, declining by 23\% from 0.26 GiB/s for 1 node to 0.20 GiB/s for 200 nodes, indicating poor scalability. Conversely, the Dardel CPU shows improved performance, with write throughput increasing from 0.09 GiB/s for 1 node to 0.41 GiB/s for 200 nodes, suggesting better suitability for parallelism. However, the Vega CPU demonstrates inconsistent performance, lacking clear scaling behavior. Given that Dardel displays the highest Write Throughput on larger nodes, we will continue our investigation using the Dardel CPU LFS, which is also applicable on the other two systems (Discoverer and Vega CPU LFS).

\begin{figure}[htbp]
    \begin{center}
        \includegraphics[width=0.95\linewidth]{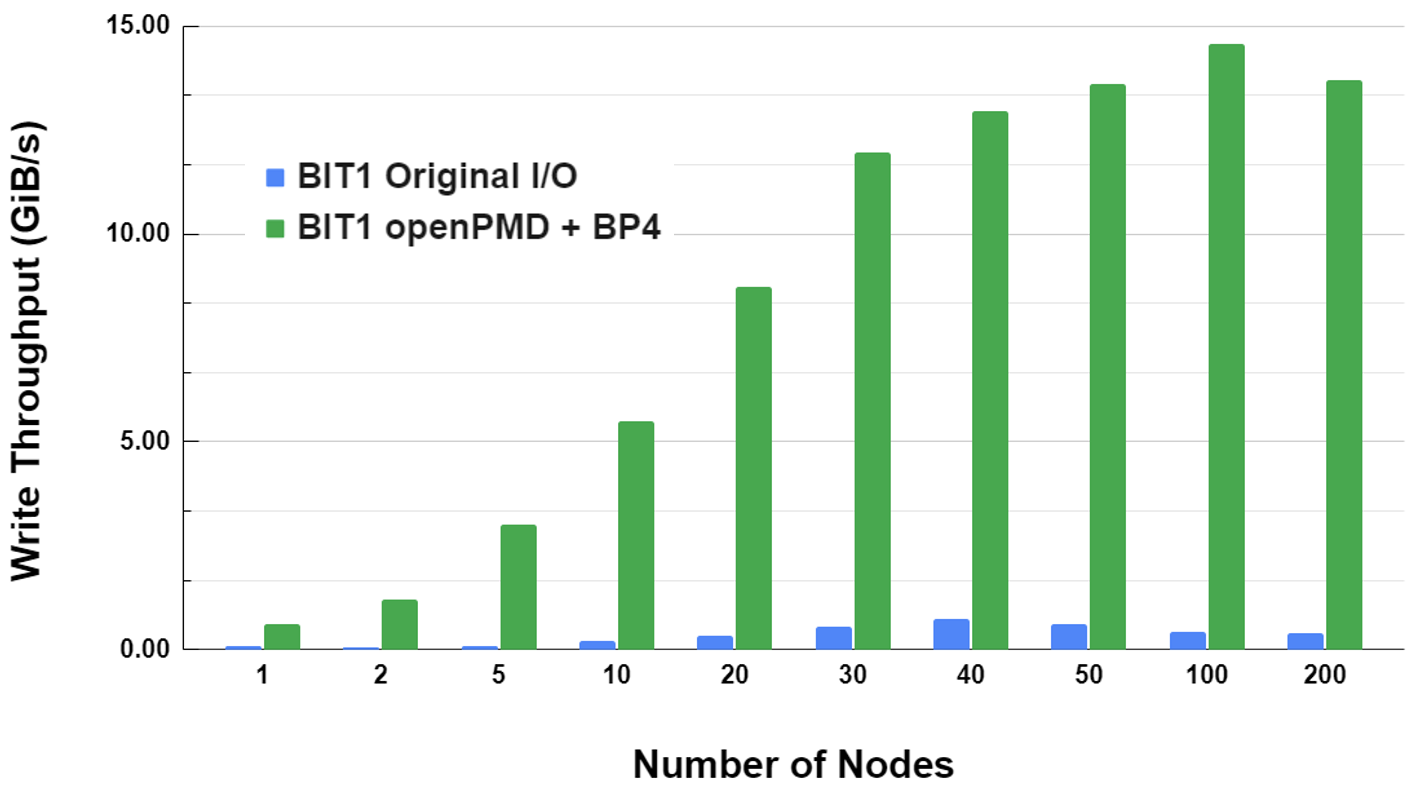}
        \caption{BIT1 Original File I/O and openPMD + BP4 Write Throughput on Dardel up to 200 Nodes, measured in GiB/s.} \label{darshan_BIT1_IO_Before_and_After_Dardel}
    \end{center}
\end{figure}

In Fig.~\ref{darshan_BIT1_IO_Before_and_After_Dardel} BIT1 Original I/O Write Throughput up to 200 nodes, performs serial I/O. The write throughput increases for small runs until the peak throughput is reached. For large runs, the I/O write throughput decreases as the cost associated with metadata write increases~\cite{williams2023leveraging}. Unlike BIT1's traditional file I/O, where the write throughput decreases primarily due to the escalating cost associated with metadata write, BIT1 openPMD + BP4 maintains a more stable throughput. This is attributed to the parallel I/O strategy applied to distribute the workload efficiently across multiple nodes, mitigating the adverse effects of metadata write overhead.





\subsection{Benchmarking}
Benchmarking, a crucial process in performance evaluation, efficiency assessment, and quality comparison, involves comparing systems, devices, or processes against established benchmarks. The IOR benchmark is a configurable tool that can be tailored to simulate the read and write operations of real-world applications. It provides options for customizing various aspects such as I/O protocols, modes, and file sizes. Essentially, IOR facilitates data reading and writing from either an exclusive or shared file. 

Below are relevant IOR parameters~\cite{iorruntimeoptions}:
\begin{itemize}[leftmargin=*]
    \item \textbf{-N (NumTasks)}:  Specifies the task count for the IOR benchmark.
    \item \textbf{-a (Api)}: Determines the API option to be used for I/O [POSIX$|$MPIIO$|$HDF5$|$\ldots$|$RADOS].
    \item \textbf{-F (FilePerProc)}: Enables file-per-process mode.
    \item \textbf{-C (ReorderTasksConstant)}: Switches task ordering to n+1 for readback.
    \item \textbf{-e (Fsync)}: Execute ``fsync" when closing POSIX write operations.
\end{itemize} 
Using the IOR benchmark, we evaluate BIT1's performance in the Dardel CPU LFS system, aiming to identify areas for improvement.

\begin{table}[htbp]
    \centering
    \begin{tabular}{|l|}
        \hline
        \begin{tabular}{@{}l@{}}
            \texttt{IOR Benchmark (FilePerProc):} \\
            \fbox{\texttt{\$ srun -n 25600 ior -N=25600 -a POSIX -F -C -e}}
        \end{tabular} \\
        \hline
        \begin{tabular}{@{}l@{}}
            \texttt{IOR Benchmark (Shared):} \\
            \fbox{\texttt{\$ srun -n 25600 ior -N=25600 -a POSIX -C -e}}
        \end{tabular} \\
        \hline
    \end{tabular}
    \vspace{2mm}
    \caption{Command lines used to run IOR~\cite{iorruntimeoptions,petersen2013optimizing} with parameters, -N, -a, -F, -C, and -e on Dardel LFS (200 nodes).}
    \label{tab:IOR_Benchmark_Commands}
\end{table}

\begin{figure}[htbp]
    \begin{center}
        \includegraphics[width=0.95\linewidth]{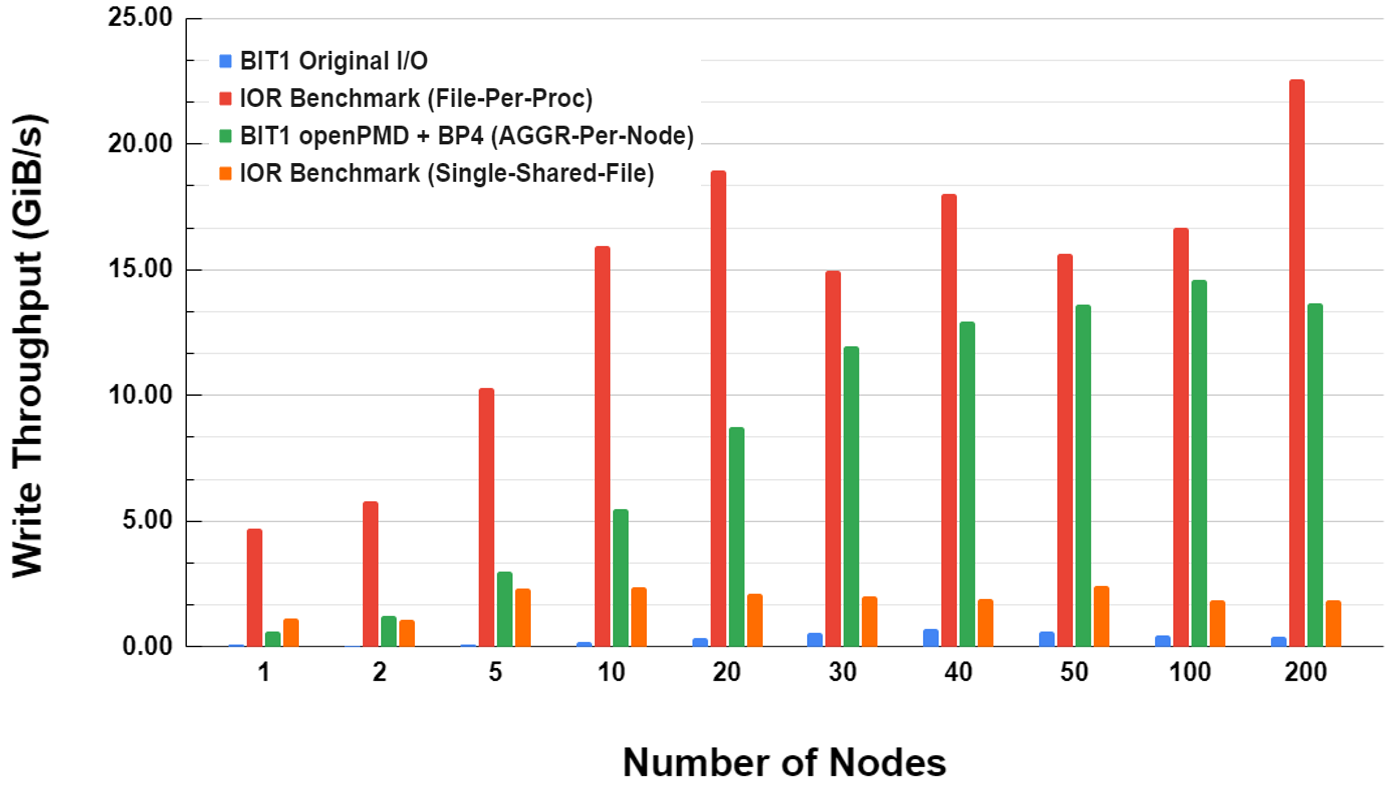}
        \caption{BIT1 I/O Write Throughput on Dardel up to 200 Nodes, measured in GiB/s.} \label{darshan_IO_Scaling-POSIX_Layer}
    \end{center}
\end{figure}

Fig. \ref{darshan_IO_Scaling-POSIX_Layer} displays the I/O performance results of the BIT1 Original I/O and BIT1 openPMD + BP4 configurations, compared to the IOR benchmarking results, using the commands outlined in Table \ref{tab:IOR_Benchmark_Commands}, on the Dardel CPU LFS system, for up to 200 nodes. Analysis reveals distinct performance trends between the two BIT1 configurations. BIT1 Original I/O exhibits a low initial write throughput of 0.09 GiB/s, gradually increasing with node count but failing to achieve competitive levels compared to the IOR benchmarks. Conversely, BIT1 openPMD + BP4 with aggregation demonstrates superior performance, starting with a higher write throughput of 0.6 GiB/s and exhibiting a notably steeper increase with additional nodes. This consistent improvement highlights the configuration's enhanced scalability, parallelization capabilities, and effectiveness in minimizing waiting times for shared I/O resources during BIT1 simulations.

\subsection{I/O Costs Per Process}


Reducing I/O function runtime is essential for enhancing the efficiency, scalability, and cost-effectiveness of computational BIT1 simulations. 
By minimizing the time spent on BIT1 I/O operations, we can accelerate data processing, optimize resource utilization, and facilitate the handling of larger datasets and more complex simulations~\cite{carns200924,logan2020extending}. 

\begin{figure}[htbp]
    \begin{center}
        \includegraphics[width=0.95\linewidth]{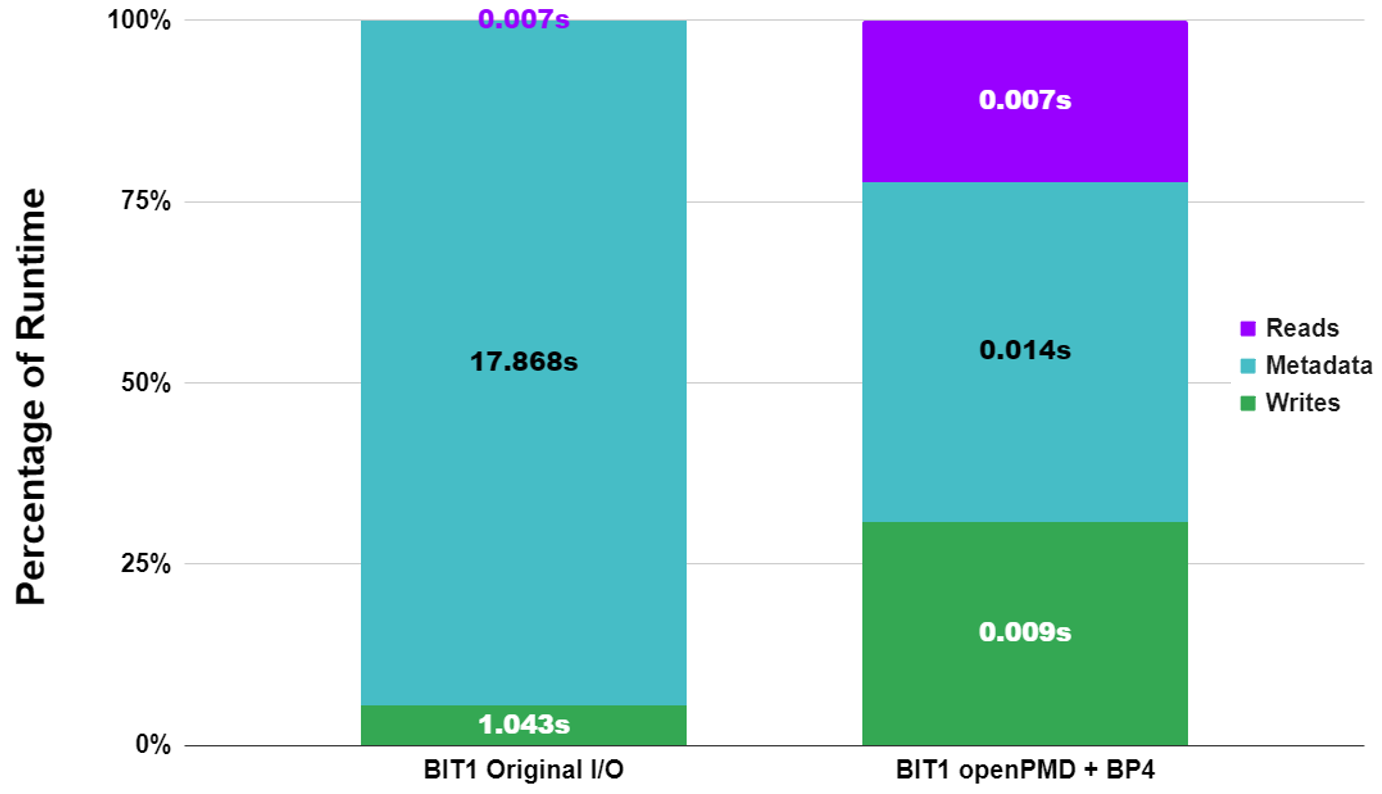}
        \caption{BIT1 Average I/O Cost Per Process on Dardel for Reads, Metadata and Writes on 200 nodes, normalized.} 
        \label{BIT1_Runtime_Percentage_Time_Spent}
    \end{center}
\end{figure}

As discovered in~\cite{williams2023leveraging, williams2024optimizing}, the peak I/O write throughput depends on the problem size, and after the peak I/O is reached, the I/O performance degrades as the metadata writing cost increases on large runs. To understand the benefits of using openPMD with ADIOS2 backends, we investigate the time spent in I/O functions on 200 nodes, comparing BIT1's original I/O method with openPMD using BP4. Figure~\ref{BIT1_Runtime_Percentage_Time_Spent} displays the normalized results of average I/O reads, writes, and metadata costs per process during BIT1 simulations on 200 nodes.

Analyzing the results reveals that the integration of openPMD with ADIOS2 backends, particularly employing BP4, has yielded remarkable enhancements in BIT1. Foremost among these improvements is the substantial reduction in metadata overhead. Previously, the average time spent on metadata operations per process stood at 17.868 seconds in the BIT1 Original I/O simulation. However, with openPMD + BP4, this time plummeted to a mere 0.014 seconds per process, representing an astounding reduction of approximately 99.92\%. Concurrently, there has been a notable enhancement in write capability. In the BIT1 Original I/O execution, the average time spent on write operations per process was 1.043 seconds, which significantly decreased to 0.009 seconds with openPMD + BP4, highlighting a reduction of around 99.14\%. 

Importantly, the time spent on reads remains consistent, primarily due to checkpointing, where files are saved and stored for restarting the simulation. This consistency shows the reliability and stability of the read operations, enabling seamless continuation of simulations while focusing on the significant improvements in metadata handling and write performance. These improvements directly contribute to overall I/O performance enhancement by efficiently managing metadata and improving write performance. As a result, BIT1 simulations benefit from the effectiveness of using openPMD with ADIOS2 backends in balancing metadata reduction and writing capability, thereby directly contributing to overall I/O performance enhancement by efficiently managing metadata and improving write performance.



\subsection{Aggregation}


For optimal I/O performance in BIT1, ``N" processes must distribute their output across ``M" files to maximize the file system's throughput, minimize the overhead of multiple processes writing to a single file, and prevent overloading the file system with excessive files. In ADIOS2, each node is allocated (or fixed) one aggregator (AGGR), leading to a single shared file among MPI processes per node~\cite{godoy2020adios}. 
\begin{figure}[htbp]
    \begin{center}
        \includegraphics[width=0.90\linewidth]{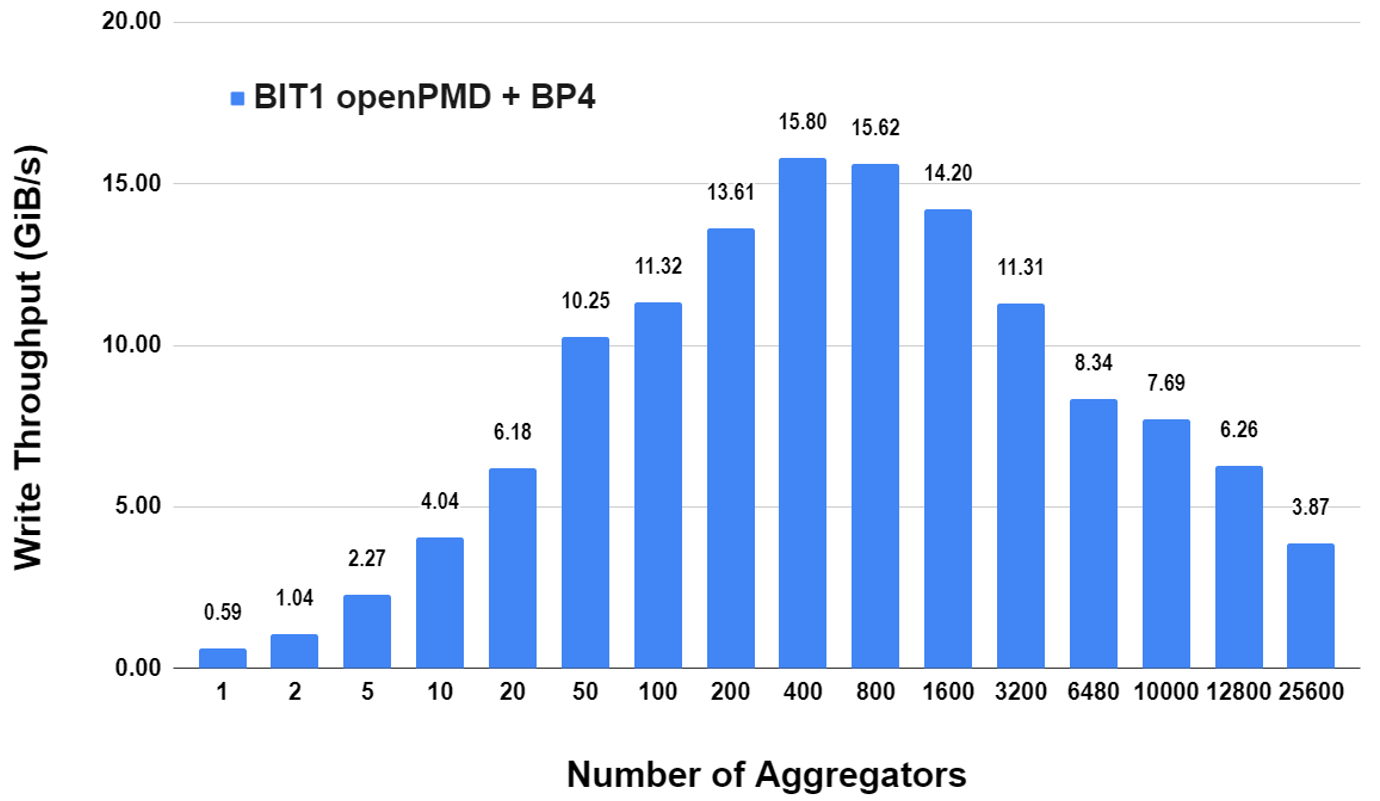}
        \caption{BIT1 openPMD + BP4 I/O Write Throughput and Aggregators on Dardel, measured on 200 nodes, in GiB/s.} 
        \label{darshan_IO_Aggregators_Scaling}
    \end{center}
\end{figure}
An essential parameter (also used with the BP5 engine) in this context is ``OPENPMD\_ADIOS2\_BP5\_NumAgg," which dictates the desired number of output files (aggregators) to be written to disk, ensuring efficient I/O performance even as the number of MPI processes varies. Fig.~\ref{darshan_IO_Aggregators_Scaling} displays the scaling results obtained from multiple BIT1 openPMD simulations using the BP4 engine, with a fixed value set for ``OPENPMD\_ADIOS2\_BP5\_NumAgg" to determine the optimal number of aggregators. Notably, as the number of aggregators increases, there is a consistent improvement in write throughput until reaching a peak at 400 aggregators (equivalent to two aggregators per node), achieving 15.80 GiB/s. Beyond this point, there is a slight decline in throughput with further aggregation, even though at the highest tested aggregation (25600), the write throughput remains significantly higher than the starting point, at 3.87 GiB/s. This demonstrates a notable increase in write throughput, highlighting the scalability benefits from 0.59 GiB/s for 1 aggregator to 15.80 GiB/s for 400 aggregators. Moreover, at 25600 aggregators, the throughput notably surpasses BIT1 Original I/O performance (0.41 GiB/s) with the same number of files, highlighting the effectiveness of the BIT1 openPMD + BP4 approach in enhancing write operation performance under extreme aggregation scenarios.

\begin{figure}[htbp]
    \begin{center}
        \includegraphics[width=0.90\linewidth]{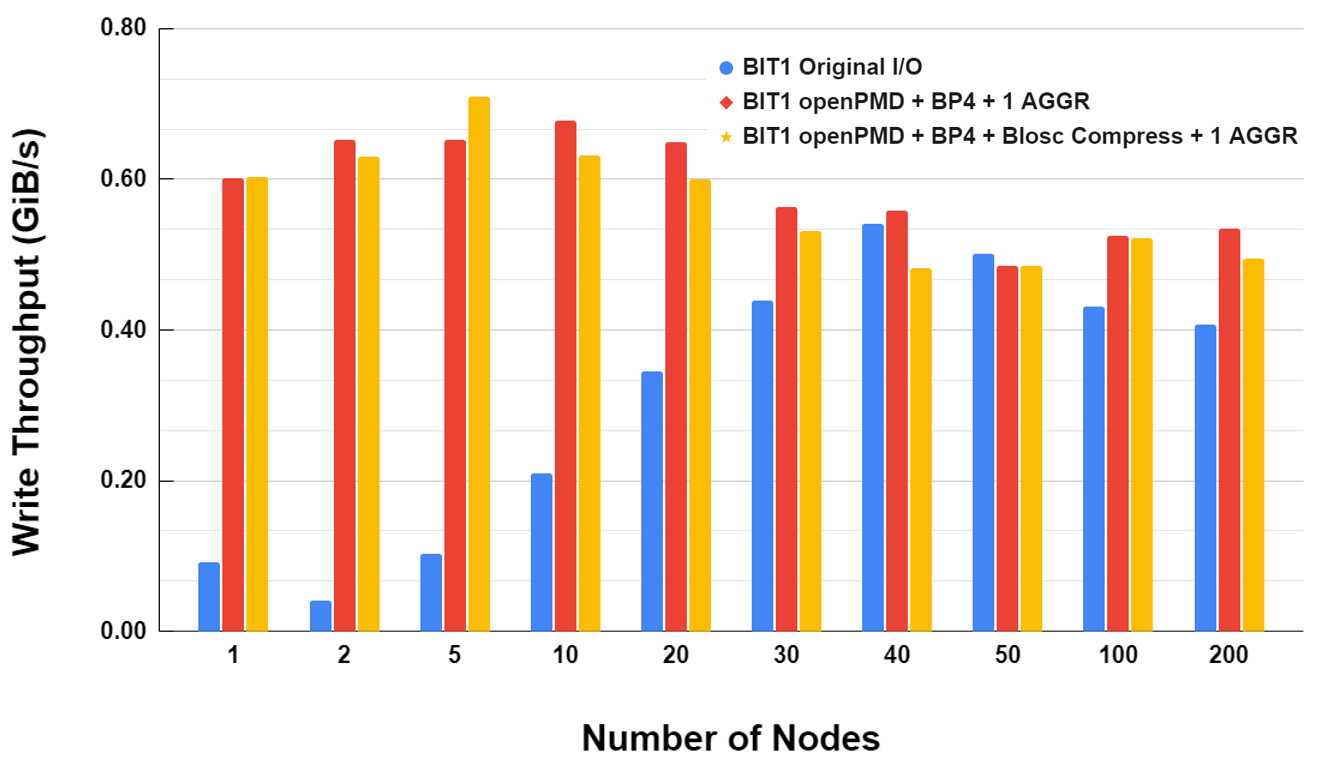}
        \caption{BIT1 I/O Write Throughput on Dardel for up to 200 Nodes, using the Blosc Compressor and one Aggregator, measured in GiB/s} \label{darshan_IO_Aggregators_Scaling_Blosc_Compressor}
    \end{center}
\end{figure}

\subsection{Data Compression}
In the BIT1 openPMD implementation and during ADIOS2 compilation, the Blosc and bzip2 compressors were chosen and enabled to enhance data transmission speed, reduce data size, and enable greater storage capacity in the file system ~\cite{pugmire2022adaptable}, all while retaining essential information for each simulation. Fig~\ref{darshan_IO_Aggregators_Scaling_Blosc_Compressor} shows scaling results obtained from BIT1 I/O Write Throughput up to 200 Nodes with Blosc compression, and one Aggregator. As before, BIT1 Original I/O displays an inconsistent performance pattern as the number of nodes increases, eventually leading to a peak write throughput of approximately 0.54 GiB/s with 40 nodes, suggesting potential inefficiencies in utilizing computational resources at higher node counts. In contrast, both ``BIT1 openPMD + BP4" configurations demonstrate enhanced scalability and efficiency, with improved performance and higher throughput observed from 1 to 10 nodes. Although compression and aggregation enhance data storage efficiency, they also introduce overhead, resulting in slightly reduced performance compared to the uncompressed configuration (BIT1 Original I/O) at higher node counts, which can be seen from 10 to 50 nodes.

Controlling ADIOS2 I/O behavior at runtime helps fine-tune our BIT1 simulation for large runs and shows greater potential for optimizing data management and processing~\cite{poeschel2021transitioning,pugmire2022adaptable}. To do this, the environment variable ``OPENPMD\_ADIOS2\_HAVE\_PROFILING" is set to ``1", providing a file (profiling.json) with profiling information at the end of each BIT1 simulation run. 

\begin{figure}[htbp]
\vspace{-0.2cm} 
    \begin{center}
        \includegraphics[width=0.90\linewidth]{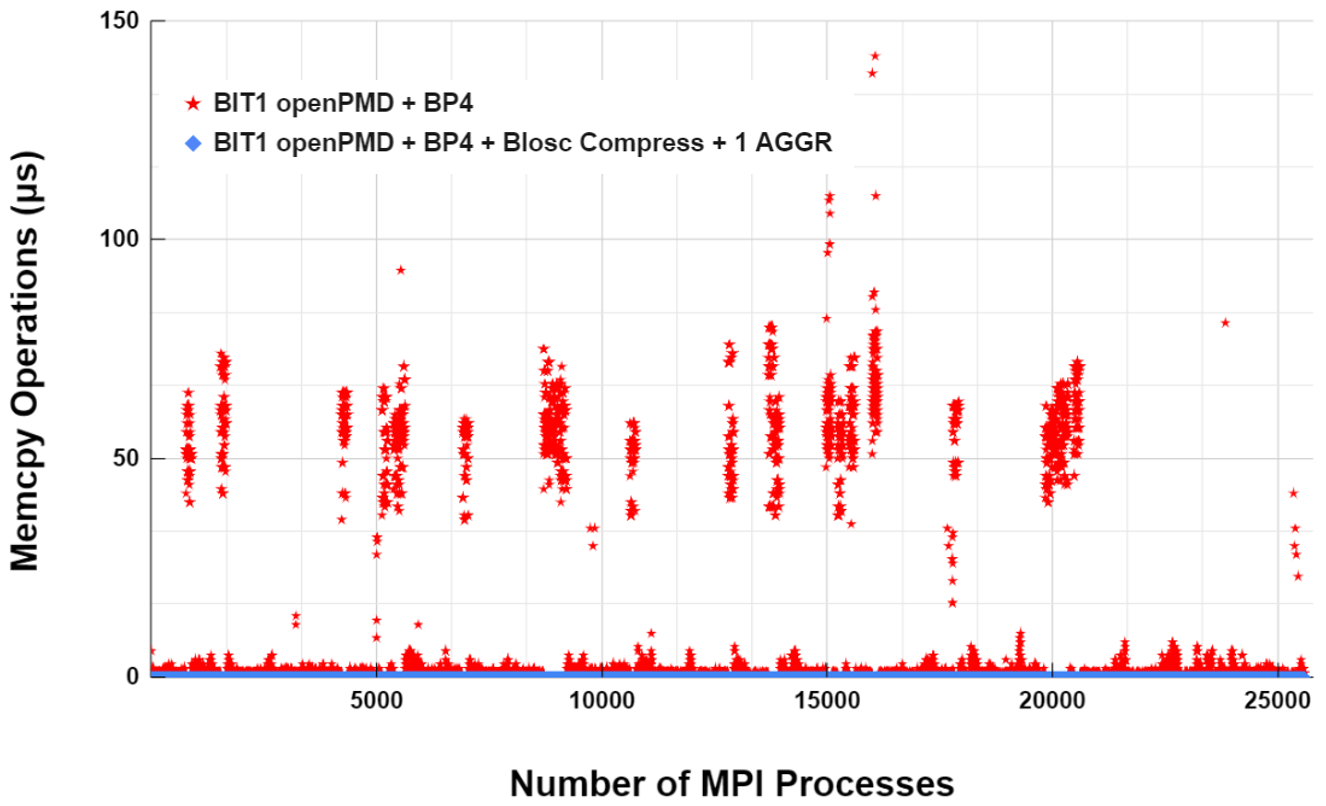}
        \caption{BIT1 openPMD + BP4 profiling.json results from Dardel on 200 nodes, showing memory copy operations executed in microseconds, with compression (blue) and without compression (red).} \label{BIT_openPMD_json_BP4_memcpy}
     \end{center}
\vspace{-0.2cm} 
\end{figure}

Fig~\ref{BIT_openPMD_json_BP4_memcpy} displays profiling.json results on 200 nodes, where memory copy operation execution times are entirely eliminated for the BIT1 openPMD + BP4 configuration with Blosc compression and 1 AGGR. This indicates that the use of Blosc compression and 1 AGGR has effectively optimized (or streamlined) the data handling process to the extent that the time spent on memory copy operations is virtually eliminated.

Table~\ref{tab:bit1_write_files_Count} displays the summary of write file scaling results for BIT1 Original without compression and BIT1 openPMD + BP4 using Blosc and Bzip2 compressors, all with varying numbers of aggregators. Analysis of the results reveals notable trends in file count and size across different configurations and node counts. In the BIT1 Original I/O setup with no compression and aggregation, the total number of written files increases significantly with the number of nodes, from 262 files for 1 node to 51,206 files with 200 nodes, while the average file size decreases as the number of nodes increases, from 1.9 MiB with 1 node to 13 KiB with 200 nodes, with a similar trend observed for maximum file size. Using compression, particularly the bzip2 compressor, alongside the openPMD + BP4 setup, leads to a substantial reduction in the total number of written files. Additionally, larger average file sizes are observed compared to the uncompressed BIT1 simulations, contributing to improved storage efficiency. For instance, on 200 nodes, the total written files decrease significantly from 6 files with 1 node to 205 files, with average file sizes ranging from 9.4 MiB to 81 MiB, and maximum file sizes varying widely from 476 MiB to 1.1 GiB. 

Shared file strategies with 1 AGGR maintain a consistent total file count across all node counts, accompanied by increased average and maximum file sizes compared to using standard setups. For instance, in the configuration using bzip2 compression and 1 AGGR, total written files remain constant at 6 files across all node counts, with average file sizes increasing slightly from 81 MiB with 1 node to 326 MiB with 200 nodes, and maximum file sizes also increasing from 476 MiB to 1.1 GiB. This is achieved by setting the ``OPENPMD\_ADIOS2\_BP5\_NumAgg" parameter to 1, ensuring that exactly one file is written on the disk for all MPI ranks. However, with the change in compression technique, particularly the Blosc compressor, alongside the openPMD + BP4 with 1 AGGR, total written files still remain constant at 6 files across all node counts with average file sizes decreasing slightly from 81 to 72 MiB for 1 node, reflecting an 11.11\% reduction, and from 326 to 314 MiB for 200 nodes, indicating a 3.68\% reduction on large runs. Additionally, maximum file sizes also decrease slightly from 476 to 422 MiB, representing an 11.43\% reduction, with roughly no change observed at 1.1 GiB for 200 nodes.

\begin{table}[htbp]
    \centering
    \resizebox{1\linewidth}{!}{%
    \begin{tabular}{||c||c|c|c|c|c|c|c|c|c|c||}
        \hline
        \multirow{2}{*}{\textbf{Number of Nodes}} & \multicolumn{10}{c||}{\rule{0pt}{2.5ex}\textbf{BIT1 Original I/O}} \\
        \cline{2-11}
        \rule{0pt}{2.5ex} & 1 & 2 & 5 & 10 & 20 & 30 & 40 & 50 & 100 & 200 \\
        \hline 
        \rule{0pt}{2.5ex} \textbf{Total Written Files} & 262 & 518 & 1286 & 2566 & 5126 & 7686 & 10246 & 12806 & 25606 & 51206\\
        \textbf{Average File Size} & 1.9MiB & 939KiB & 381KiB & 192KiB & 98KiB & 67KiB & 51KiB & 41KiB & 22KiB & 13KiB \\
        \textbf{Max File Size} & 3.8MiB & 1.9MiB & 763KiB & 383KiB & 194KiB & 130KiB & 98KiB & 79KiB & 40KiB & 25KiB \\
        \hline
        \multicolumn{11}{c}{} \\
        \hline
        \multirow{2}{*}{\textbf{Number of Nodes}} & \multicolumn{10}{c||}{\rule{0pt}{2.5ex}\textbf{BIT1 openPMD + BP4}} \\
        \cline{2-11} 
        \rule{0pt}{2.5ex} & 1 & 2 & 5 & 10 & 20 & 30 & 40 & 50 & 100 & 200 \\
        \hline
         \rule{0pt}{2.5ex} \textbf{Total Written Files} & 6 & 7 & 10 & 15 & 25 & 35 & 45 & 55 & 105 & 205 \\
        \textbf{Average File Size} & 81MiB & 70MiB & 51MiB & 37MiB & 25MiB & 20MiB & 17MiB & 16MiB & 12MiB & 9.4MiB \\
        \textbf{Max File Size} & 476MiB & 239MiB & 97MiB & 53MiB & 106MiB & 158MiB & 211MiB & 263MiB & 526MiB & 1.1GiB \\
        \hline
        \multicolumn{11}{c}{} \\
        \hline
        \multirow{2}{*}{\textbf{Number of Nodes}} & \multicolumn{10}{c||}{\rule{0pt}{2.5ex}\textbf{BIT1 openPMD + BP4 + 1 AGGR}} \\
        \cline{2-11}
        \rule{0pt}{2.5ex} & 1 & 2 & 5 & 10 & 20 & 30 & 40 & 50 & 100 & 200 \\
        \hline
        \rule{0pt}{2.5ex} \textbf{Total Written Files} & 6 & 6 & 6 & 6 & 6 & 6 & 6 & 6 & 6 & 6 \\
        \textbf{Average File Size} & 81MiB & 82MiB & 86MiB & 92MiB & 104MiB & 116MiB & 128MiB & 140MiB & 202MiB & 326MiB \\
        \textbf{Max File Size} & 476MiB & 478MiB & 484MiB & 493MiB & 511MiB & 529MiB & 548MiB & 567MiB & 665MiB & 1.1GiB \\
        \hline
        \multicolumn{11}{c}{} \\
        \hline
        \multirow{2}{*}{\textbf{Number of Nodes}} & \multicolumn{10}{c||}{\rule{0pt}{2.5ex}\textbf{BIT1 openPMD + BP4 + Blosc Compress + 1 AGGR}} \\
        \cline{2-11} 
        \rule{0pt}{2.5ex} & 1 & 2 & 5 & 10 & 20 & 30 & 40 & 50 & 100 & 200 \\
        \hline
       \rule{0pt}{2.5ex} \textbf{Total Written Files} & 6 & 6 & 6 & 6 & 6 & 6 & 6 & 6 & 6 & 6 \\
        \textbf{Average File Size} & 72MiB & 73MiB & 76MiB & 83MiB & 95MiB & 107MiB & 119MiB & 131MiB & 192MiB & 314MiB \\
        \textbf{Max File Size} & 422MiB & 424MiB & 429MiB & 437MiB & 456MiB & 473MiB & 490MiB & 506MiB & 590MiB & 1.1GiB \\
        \hline
    \end{tabular}
    }
    \vspace{2mm}
    \caption{BIT1 Writes Files on Dardel CPU LFS for up to 200 nodes across different configurations, showing the total number of files created along with their average and maximum sizes (in MiB).}
     \label{tab:bit1_write_files_Count}
\end{table}


\subsection{Lustre File Striping}

Lustre file striping plays a crucial role in optimizing BIT1 storage performance. When a file is written to Lustre, it is divided into smaller segments called "stripes". These stripes are then systematically distributed across multiple OSTs. By separating the data across several OSTs, Lustre can leverage the aggregate throughput and I/O capabilities of these devices, enabling faster I/O operations~\cite{lin2013optimizing, yu2008performance}.

The selection of Lustre file striping configuration, which includes parameters such as the number of OSTs and the size of the stripes, has a direct impact on system performance. A well-tailored striping configuration can optimize throughput and reduce latency by efficiently distributing the workload across available storage resources~\cite{markidis2021understanding, saini2012performance, yu2008performance}. 


\begin{table}[htbp]
    \centering
    \begin{tabular}{|c|}
        \hline
        Lustre File Striping Command \\
        \hline
        \fbox{\texttt{lfs setstripe -c 8 -S 16M io\_openPMD}} \\
        \hline
    \end{tabular}
    \vspace{2mm}
    \caption{Command line to configure Dardel's LFS to stripe newly created files within the ``io\_openPMD" directory across 8 OSTs, with each stripe size set to 16,777,216 bytes (16 MiB).}
    \label{tab:lustre_file_striping_command}
\end{table}

The command line shown in Table~\ref{tab:lustre_file_striping_command} is used to configure Lustre file striping and set up Dardel’s Lustre parallel file system. The command ``lfs setstripe" determines how files are distributed across multiple OSTs. In this command, ``-c 8" specifies that each file will be divided into 8 parts (stripes), while ``-S 16M" indicates that each stripe will have a size of 16 MiB during creation in the ``io\_openPMD" directory. To verify that the file striping configuration has been successfully applied, ``lfs getstripe" command is used to extract this information. The output displayed in Listing~\ref{5th:lfs_getstripe_output} provides essential details about the file striping configuration. It reveals that the file ``data.0" in the directory ``io\_openPMD/dat\_file.bp4" is divided into 16 stripes, each with a size of 16,777,216 bytes (16 MiB), utilizing the raid0 striping pattern. Further details provided include the object directory index, object ID, associated group, and the utilization of 8 OSTs for file striping.

\vspace{2mm}

\begin{lstlisting}[style=custom,
caption={Extracted Lustre file striping configuration for file ``data.0" using the ``lfs getstripe'' command. The file is divided into 8 pieces (stripes), each with a size of 16,777,216 bytes (16 MiB), utilizing the raid0 (round-robin fashion) striping pattern. Detailed information such as Object Directory index, object ID, associated group, and the use of 8 OSTs for file striping.}, label={5th:lfs_getstripe_output}]
$ lfs getstripe io_openPMD/dat_file.bp4/data.0
io_openPMD/dat_file.bp4/data.0
lmm_stripe_count:  8
lmm_stripe_size:   16777216
lmm_pattern:       raid0
lmm_layout_gen:    0
lmm_stripe_offset: 17
        obdidx           objid           objid           group
            17       297315680     0x11b8ad60      0x700000400
            19       297401760     0x11b9fda0      0x740000400
            21       297299648     0x11b86ec0      0x800000400
            23       297230944     0x11b76260      0x840000400
            25       296891424     0x11b23420      0x900000400
            27       297129552     0x11b5d650      0x940000400
            29       294976177     0x1194fab1      0xa00000400
            31       297343489     0x11b91a01      0xa40000400
\end{lstlisting}

\begin{figure}[htbp]
    \begin{center}
        \includegraphics[width=0.95\linewidth]{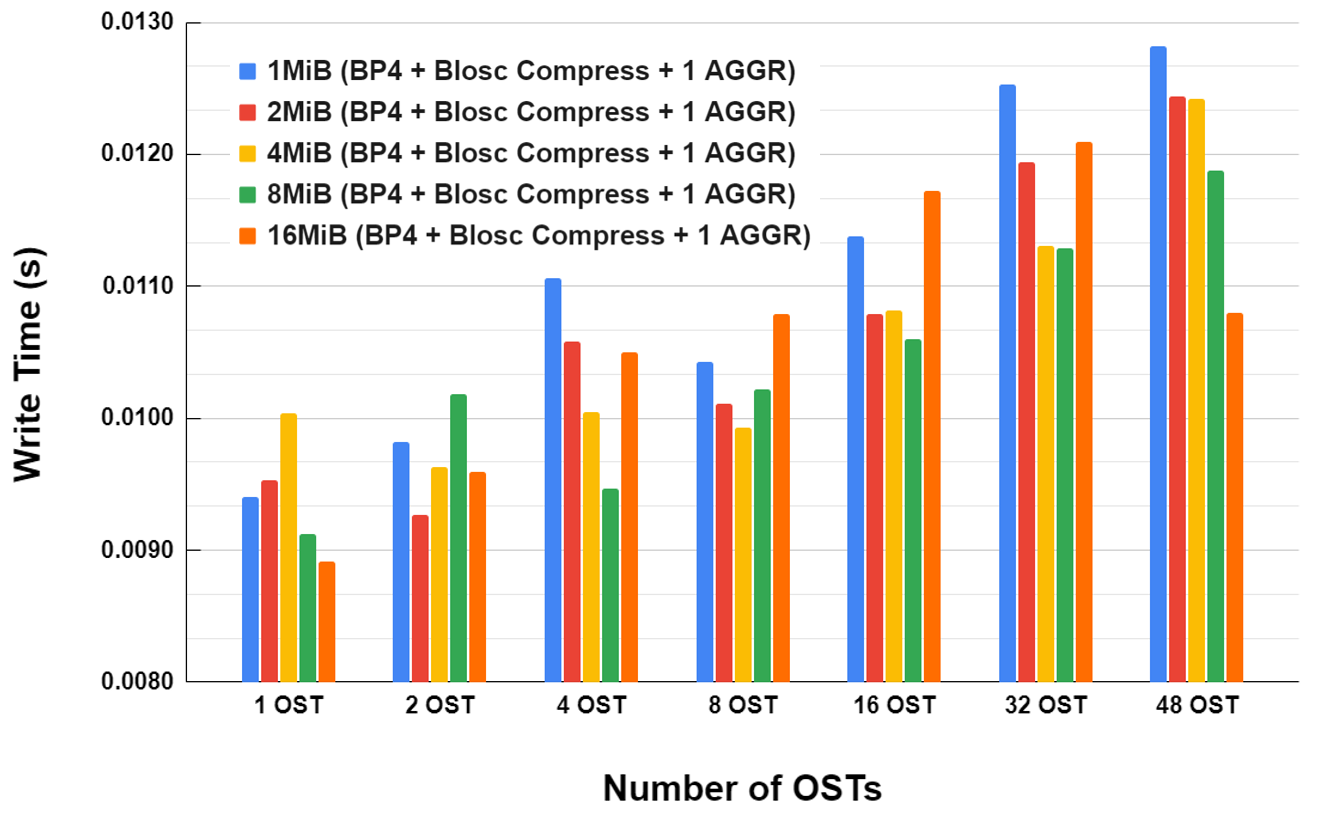}
        \caption{BIT1 openPMD + BP4 I/O Write Time Spent on Dardel CPU LFS (200 nodes) using Lustre File Striping with Blosc Compressor and one Aggregator, measured in seconds.} \label{darshan_IO_OSTs_Aggregator_Scaling_Blosc_Compressor}
    \end{center}
\end{figure}

Focusing on evaluating the write time spent and determining the optimal Lustre configuration across different Lustre stripe sizes and varying the number of OSTs, Figure~\ref{darshan_IO_OSTs_Aggregator_Scaling_Blosc_Compressor} offers insights into the direct impact of system performance for BIT1 openPMD + BP4 configurations utilizing Lustre file striping, Blosc compression, and one AGGR on 200 nodes. Analyzing the scaling results reveals interesting findings. Smaller Lustre stripe sizes tend to yield better performance, particularly noticeable when employing a single OST, with optimal performance achieved at 0.0089s with a Lustre stripe size of 16MiB. However, the relationship between Lustre stripe size and write time varies significantly based on the number of OSTs employed. Interestingly, in some scenarios, increasing the number of OSTs led to reduced write times, indicating potential improvements in workload distribution and parallelism. For instance, with a Lustre stripe size of 4MiB, the write time decreases by approximately 4\% when transitioning from 1 OST to 2 OSTs. Conversely, with a stripe size of 16MiB, the write time increases by approximately 7.87\% with the same transition. Nonetheless, these trends are not uniform across all configurations, highlighting the need for tailored optimization strategies in Dardel CPU LFS. 

Moreover, diminishing returns were observed beyond a certain point, where further increasing OSTs provided minimal or negligible improvements in write time. Finding an optimal balance is essential to avoid unnecessary overhead. This suggests the importance of carefully fine-tuning Lustre configurations, especially with smaller stripe sizes, as a smaller number of OSTs may lead to more efficient performance for write operations in the BIT1 openPMD + BP4 configurations on large runs.


\section{Related Work} \label{sec:relwork}
PIC codes are fundamental for plasma simulations, with significant efforts in development, analysis, optimization, and integration of modern standards like openPMD and ADIOS2. The use of openPMD and ADIOS2 has been shown to improve I/O performance. By utilizing the ADIOS2 library through the openPMD-api, performance numbers can be collected, and models can be built to enhance I/O performance~\cite{wan2021improving}. ADIOS2 provides a framework for high-performance I/O, expanding on the legacy of earlier versions and offering C++ and Python APIs for scientific I/O with openPMD~\cite{pugmire2022adaptable, godoy2020adios}. It also offers an abstraction for high-performance I/O and supports features such as data staging and compression, leading to improved I/O workflows and data reduction~\cite {huebl2017scalability}. Banerjee et al.~\cite{banerjee2022algorithmic} further highlighted an algorithmic and software pipeline for data compression with error guarantees, utilizing ADIOS2 for efficient parallel I/O workloads. Research has also shown that organizing large datasets efficiently can lead to improved analyses on HPC systems, further optimize I/O performance~\cite{gu2022organizing}. The concept of improving I/O performance has been a topic of interest in the field of computer science for several decades. Crockett~\cite{crockett1989file} discussed file concepts for parallel I/O, highlighting the importance of efficient data storage and retrieval in parallel computing environments. Similarly, James et al.~\cite{james1990distributed} introduced a distributed-directory scheme to achieve a scalable and coherent interface in parallel systems. Pioneering research by Berkeley University and Lawrence Livermore National Laboratory has laid the foundation for understanding PIC methods and their implementations, as documented in influential works by Birdsall and Langdon~\cite{birdsall2018plasma, birdsall1991particle}. One of the key features of the BIT1 code is its memory layout. The data layout was explained in~\cite{tskhakaya2007optimization}, and a detailed explanation of the governing equations and algorithmic part is provided in~\cite{tskhakaya2010pic}, offering insights into the foundational principles driving BIT1's computational framework. Williams et al. conducted a detailed performance analysis of BIT1, leveraging various profiling tools to unravel its computational dynamics and identify potential optimization opportunities~\cite{williams2023leveraging, williams2024optimizing}. This study sheds light on the complexities of BIT1's execution, providing valuable guidance for enhancing its performance. Recent advancements in data handling paradigms have revolutionized the I/O workflows of PIC simulations. Poeschel et al.~\cite{poeschel2021transitioning} demonstrated the transformative potential of integrating openPMD and ADIOS2 into PIConGPU, transitioning from traditional file-based approaches to streamlined data pipelines. 


\section{Discussion and Conclusion} \label{sec:conc}
Our primary goal was to enhance the write output throughput and scalability of BIT1 performance. Central to our key efforts in this work was the integration of the openPMD standard, which serves as a foundation for efficient parallel write operations. By prioritizing support for parallel workflows and emphasizing the minimization of data and storage requirements on the file system, we have uncovered significant insights. Most importantly, it's worth noting that read operations were not a bottleneck that needed addressing, as checkpoints read very little data in BIT1.

Through comprehensive benchmarking and cost analysis using the Darshan I/O performance monitoring tool, we demonstrated remarkable scalability and write throughput improvements achievable with the openPMD integration, particularly notable with the ADIOS2 BP4 engine. Our exploration of data compression and aggregation techniques has highlighted crucial roles in enhancing data storage efficiency, even though these techniques introduce minor performance trade-offs, such as overhead. Our investigation into using Lustre file striping configurations has revealed a substantial impact on system performance. Notably, we observed that optimal performance varies depending on the number of OSTs, stripe sizes, and if data compression and aggregation techniques are used, especially within the context of BIT1 openPMD + BP4 configurations on large-scale runs. 

Future research can enhance BIT1's capabilities by prioritizing openPMD integration (with ADIOS2), investigating parallel post processing performance benchmarks, particle load balancing, and continuing with checkpoint restarts towards evaluating and improving resilience capabilities. Moreover, future research should thoroughly investigate the utilization of other supported engines like the Sustainable Staging Transport (SST). The ADIOS2 SST engine enables the direct connection of data producers and consumers via the ADIOS2 write/read APIs, facilitating the movement of data between processes for in-situ processing, analysis, and visualization. These improvements would further minimize data and storage requirements, ultimately fostering more efficient and scalable BIT1 simulations.




\end{document}